\documentclass[a4paper,fleqn]{cas-sc}

\usepackage[authoryear,longnamesfirst]{natbib}

\usepackage{amsmath,amssymb,amsfonts}
\usepackage{algorithmic}
\usepackage{graphicx}
\usepackage{algorithm,algorithmic}
\usepackage{hyperref}
\hypersetup{hidelinks}
\usepackage{textcomp}
\usepackage{flushend}
\usepackage{booktabs}
\usepackage{multirow}
\usepackage{url}
\usepackage{subcaption,color}

\newcommand{\eg}{{\em e.g.}}           
\newcommand{\ie}{{\em i.e.}}           

\def\tsc#1{\csdef{#1}{\textsc{\lowercase{#1}}\xspace}}
\tsc{WGM}
\tsc{QE}
\tsc{EP}
\tsc{PMS}
\tsc{BEC}
\tsc{DE}

\begin{document}
\let\WriteBookmarks\relax
\def\floatpagepagefraction{1}
\def\textpagefraction{.001}
\shorttitle{IdenBAT: Disentangled Representation Learning for Identity-Preserved Brain Age Transformation}
\shortauthors{Maeng et~al.}

\title [mode = title]{IdenBAT: Disentangled Representation Learning for Identity-Preserved Brain Age Transformation}                      



\author[1]{{Junyeong Maeng}}
\ead{mjy8086@korea.ac.kr}

\credit{Methodology, Software, Writing - original draft, Investigation}

\affiliation[1]{organization={Department of Artificial Intelligence, Korea University},
                city={Seoul},
                postcode={02841}, 
                country={Republic of Korea}}

\affiliation[2]{organization={Department of Brain and Cognitive Engineering, Korea University},
                city={Seoul},
                postcode={02841}, 
                country={Republic of Korea}}

\author[1]{{Kwanseok Oh}}

\credit{Writing - review \& editing, Formal analysis}

\author[2]{{Wonsik Jung}}

\credit{Data curation, Writing - review \& editing}

\author[1,2]{{Heung-Il Suk}}
\cormark[1]
\ead{hisuk@korea.ac.kr}

\credit{Writing - review \& editing, Supervision, Project administration, Funding acquisition, Conceptualization}

\cortext[cor1]{Corresponding author}

\begin{abstract}
Brain age transformation aims to convert reference brain images into synthesized images that accurately reflect the age-specific features of a target age group. The primary objective of this task is to modify only the age-related attributes of the reference image while preserving all other age-irrelevant attributes. However, achieving this goal poses substantial challenges due to the inherent entanglement of various image attributes within features extracted from a backbone encoder, resulting in simultaneous alterations during the image generation. To address this challenge, we propose a novel architecture that employs disentangled representation learning for identity-preserved brain age transformation called IdenBAT. This approach facilitates the decomposition of image features, ensuring the preservation of individual traits while selectively transforming age-related characteristics to match those of the target age group. Through comprehensive experiments conducted on both 2D and full-size 3D brain datasets, our method adeptly converts input images to target age while retaining individual characteristics accurately. Furthermore, our approach demonstrates superiority over existing state-of-the-art regarding performance fidelity. Code is available at: \url{https://github.com/ku-milab/IdenBAT}.

\end{abstract}

\begin{keywords}
Brain Aging \sep Feature Disentanglement \sep Identity Preservation \sep Image-to-Image Translation \sep Magnetic Resonance Imaging
\end{keywords}

\maketitle

\section{Introduction}

Brain aging represents an intrinsic biological phenomenon marked by discernible morphological changes within the human brain~\cite{fjell2010structural}. In the analysis of brain aging using medical imaging, structural magnetic resonance imaging (sMRI) plays a crucial role as they provide detailed insights into age-related variations and assist in accurate assessments of these alterations. Advances in sMRI-based age transformation have especially allowed researchers and clinicians to visualize and quantify patient-specific intricate brain maturation and degeneration patterns, facilitating medical diagnosis advancements. These capabilities can be pivotal for longitudinal studies to track cognitive or health state progressions over time~\cite{cole2018brain,huizinga2018spatio}, whereas brain age transformation with preserving patient traits remains a formidable challenge. Because most methods even change characteristics unrelated to aging during the transformation process, the crux lies in modeling the aging process without distorting personal identities intrinsic to each subject~\cite{xia2021learning}. When the aging model fails to preserve personal properties regarding identity, it may lead to misinterpretations of age-related changes, potentially compromising the accuracy and reliability of diagnostic decisions.

\begin{figure*}[t]
    \centering
    \includegraphics[scale=0.120]{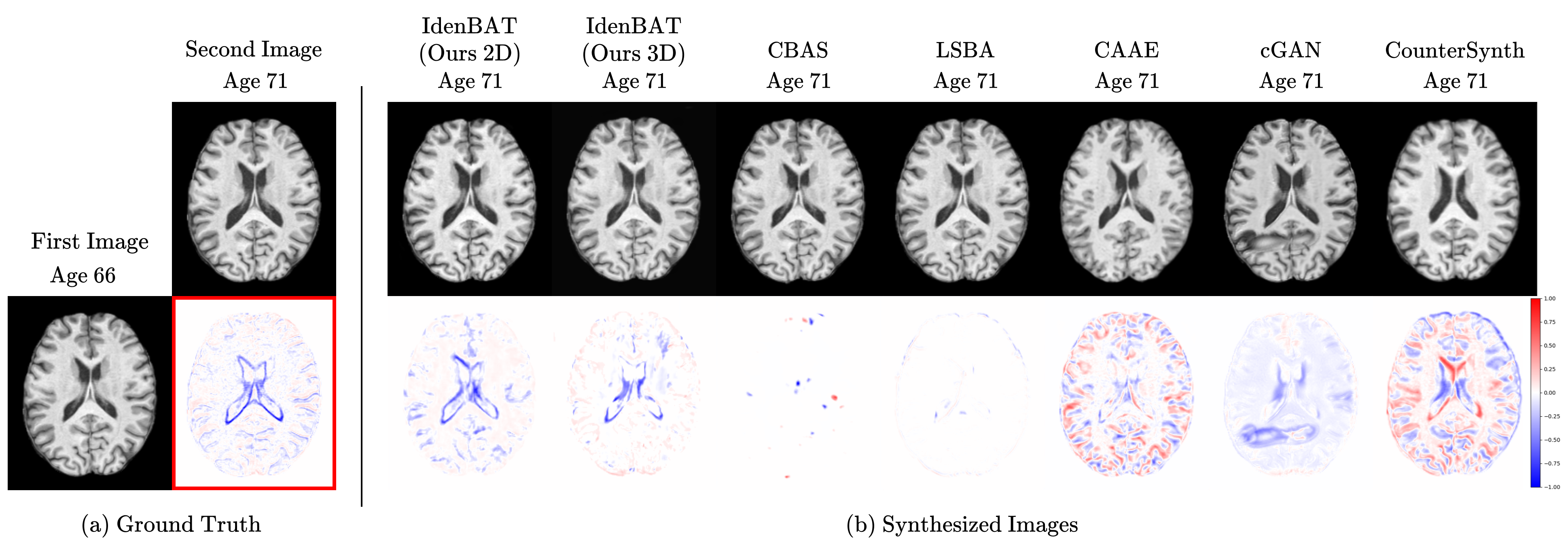}
    \caption{Visualization of results between subjects' follow-up scans (Ground Truth) and brain age transformation studies (Synthesized Images). (a) shows ground truth images of the longitudinal MRI scans and their difference map (red box). (b) displays synthesized images with age 71 and the difference maps between their and ground truth image with age 66.}\label{fig:gt}
\end{figure*}

Previous brain age transformation studies~\cite{huizinga2018spatio,zhang2016consistent,zhao2019variational,lorenzi2015disentangling,sivera2019model} have often relied on prototype-based strategies that compare averaged brain patterns across different age groups. While these approaches aid in understanding generalized characteristics shared among age groups, they tend to neglect the unique traits of individual subjects. Recently, with the emergence of generative models using longitudinal data~\cite{goodfellow2014generative,makhzani2015adversarial}, researchers have gained the ability to create more accurate and realistic simulations of brain aging by virtue of the advantages of its data, which comprised MRI scans of the same subject at multiple time points~\cite{rachmadi2019predicting,rachmadi2020automatic,wegmayr2019generative,ravi2019degenerative}. These approaches, considering the temporal properties, allow for a better understanding of the aging progression and efficiently preserve the individual identity features. However, acquiring comprehensive longitudinal data poses challenges in the medical field owing to the need for long-term follow-up and the high costs associated with multiple sMRI imaging~\cite{xia2021learning}.

To alleviate such limitations, brain aging studies~\cite{xia2019consistent,xia2021learning,pombo2023equitable} have introduced conditional generative adversarial network (cGAN)~\cite{mirza2014conditional} framework that uses cross-sectional data instead of relying on longitudinal data. By addressing age-related information during generation, cGAN-based models are proficient at age-specific transformation according to age condition. In this context, \cite{xia2021learning} utilizes adversarial learning, employing MRI scans of subjects at the target age as real images to transform input images to a different age. Similarly, \cite{pombo2023equitable} designs a discriminator that performs an age regression task, ensuring accurate conversion of generated images to the target age following the principles of StarGAN~\cite{choi2018stargan}.

However, cGAN-based models using cross-sectional data pose a limitation in image synthesis because they often fail to accurately maintain the subject’s identity~\cite{ziegler2012models}. Specifically, the shortcoming of such methods is their inability to leverage temporal information gathered from long-term follow-up of the input images. To deal with this, \cite{xia2019consistent,xia2021learning} implemented objective functions that consider structural similarity, individually yielding reconstructed outputs akin to the input reference image. Moreover, \cite{pombo2023equitable} employed the diffeomorphic spatial deformation model~\cite{dorta2020gan,krebs2018unsupervised} to preserve subject identity and target pathological signals. Nonetheless, a substantial limitation emerges from focusing solely on low-level structural information within image regions. In this light, these approaches superficially retain identity while excluding the consideration of high-level features intricately entangled with age and identity characteristics. Thus, it is necessary to disentangle identity-related features from intertwined high-level information and perform the elaborate incorporation of age-associated details into the brain aging process.

\begin{figure*}[t]
\centering
{\includegraphics[width=1.01\textwidth]{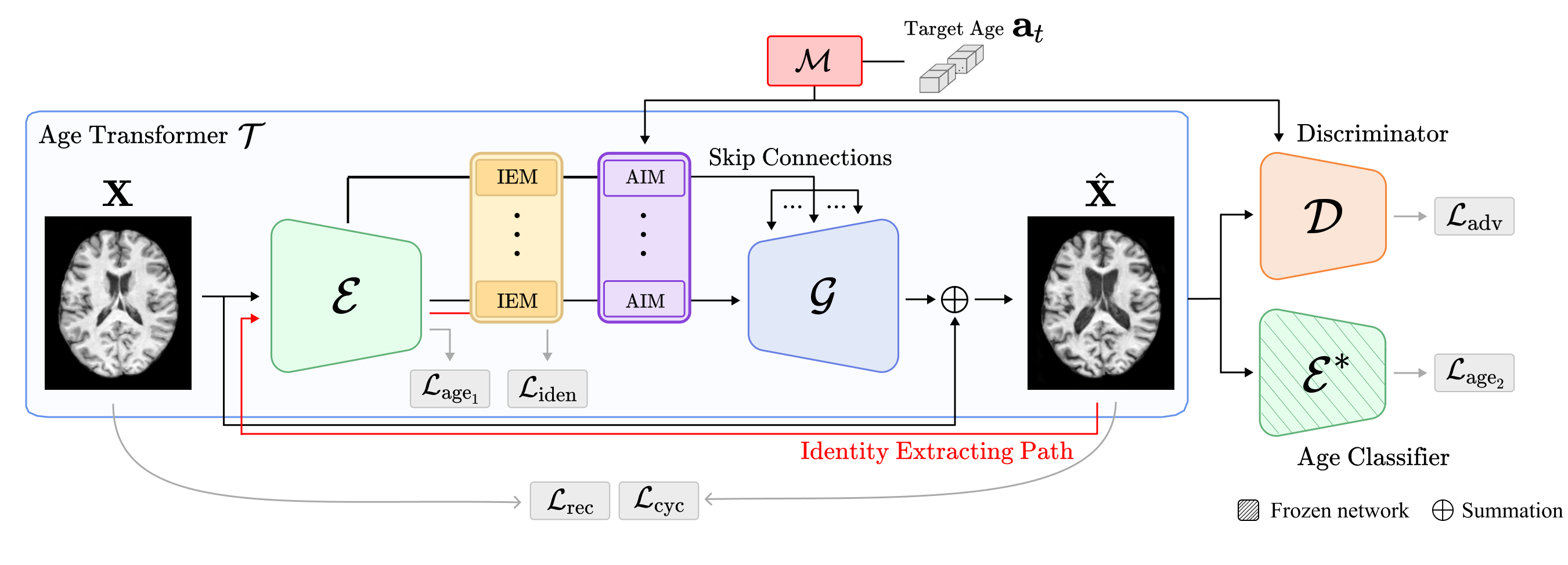}}
\vspace{-0.5cm}
\caption{Overall architecture of IdenBAT for brain age transformation. Age transformer $\mathcal{T}$, which consists of encoder $\mathcal{E}$, identity extracting module (IEM), age injecting module (AIM), and generator $\mathcal{G}$ aims to synthesize age-transformed image $\hat{\mathbf{X}}$.}
\label{fig:architecture1}
\end{figure*}

In this work, we propose disentangled representation learning for identity-preserved brain age transformation (IdenBAT). The proposed IdenBAT is based on cGAN architecture, which consists of the encoder $\mathcal{E}$, identity extracting module (IEM), age injecting module (AIM), generator $\mathcal{G}$, and discriminator $\mathcal{D}$ as illustrated in In Fig.~\ref{fig:architecture1}. Specifically, initial features extracted from the encoder are disentangled into age-related and age-irrelevant features through IEM. The age-irrelevant feature encapsulates the personal traits inherent in an input image, representing identity-related features. The age-converted image should incorporate target age information based on the original image's identity. To this end, we inject the age condition into the identity-related feature through a style transfer mechanism~\cite{karras2020analyzing} in AIM. In this way, the age and identity information is seamlessly combined for sophisticated image synthesis that maintains the identity of the reference subject while accurately reflecting the target ages.

Our approach is grounded in two fundamental premises regarding alignment within the manifold space to ensure efficient feature disentanglement. First, we assume that a reference image and its age-transformed counterpart solely differ in age-related characteristics, with all other attributes remaining constant. This hypothesis is based on the understanding that aging primarily impacts specific biological and morphological features while an individual’s fundamental identity remains consistent over time~\cite{bacon2021epigenetics}. To apply this concept, we design the input and age-converted images as two different perspectives of a single subject and maximize the similarity of their identity-related features. Second, age-related and identity-related features should exhibit orthogonality in the feature space. This orthogonal relationship ensures that alterations in age conditions do not affect other characteristics, such as the identity of the input subject, thereby preserving personal traits across different ages. This separation reflects the practical distinction between the aging progression and an individual’s inherent identity, resulting in more accurate and faithful representations in the age-transformed images~\cite{ham2020advances}. In this context, we employ a constraint to ensure zero cosine similarity between age-related and identity-related features to enforce their orthogonality. Throughout these straightforward methods of disentangled representation learning, IdenBAT effectively decomposes age-related and identity-related features from the entangled latent representation.

The primary contributions of this study are summarized as follows: (1) We propose a novel disentangled representation learning scheme to separate age and identity-related features, preserving personal traits in brain age transformation. (2) The proposed IdenBAT generates age-converted and identity-preserved images without requiring longitudinal data by directing target age information to the identity-related features. (3) We empirically demonstrated the superiority of the proposed model compared to existing state-of-the-art methods and its applicability in brain age transformation using two-dimensional (2D) and full-size three-dimensional (3D) brain images.

The proposed IdenBAT builds upon our preliminary work, GMBA~\cite{maeng2023age}, which focuses on realistic face age transformation using age-aware guidance via masking attention. The advancements of IdenBAT are as follows: (1) Different perspectives on identity preservation. While GMBA primarily focuses on preserving non-aging areas, such as the background in facial images, IdenBAT prioritizes the preservation of personal traits. To this end, IdenBAT employs advanced techniques to disentangle identity-related features from age-related features, ensuring that the unique characteristics of the individual are effectively maintained throughout the age transformation.
(2) Eliminating the need for a pre-trained identity classifier. In brain age transformation, the lack of large-scale pre-trained identity classifier~\cite{ubayashi2010archface} poses challenges for identity preservation. IdenBAT addresses this issue by introducing a novel identity-extracting loss.
(3) Available for various dimensional data. Contrary to GMBA, which is tailored to 2D facial images, exploiting IdenBAT allows identity-preserved age transformation in 2D as well as full-size 3D MRIs. Thus, this method specialized for medical images has become possible to reflect the volume and spatial characteristics of sMRI.

\section{Related Work}

\subsection{Brain Age Transformation}
Traditional brain aging studies encompass biological-based and prototype-based methods. Biological-based methods~\cite{khanal2017simulating} analyze images' biological and physical properties, such as MRI volume changes, to convert age. However, these models often require precise human knowledge or extensive longitudinal data. Conversely, prototype-based methods~\cite{zhang2016consistent,huizinga2018spatio,sivera2019model} cluster images into age groups and calculate the average prototype of each group. While effective, these approaches overlook the intrinsic attributes of individual images by relying on representative averaged images.

With the advent of deep learning, deep generative model-based methods for brain age transformation have emerged, showcasing substantial advancements surpassing traditional approaches. \cite{rachmadi2019predicting,rachmadi2020automatic,wegmayr2019generative} utilize generative adversarial network (GAN)~\cite{goodfellow2014generative}-based architectures to capture structural changes in MRI scans, while \cite{ravi2019degenerative} employs conditional adversarial autoencoder (CAAE)~\cite{zhang2017age} to synthesize aged brain images. However, these methods typically require longitudinal data for model training. \cite{zhao2019variational,pawlowski2020deep} utilize variational autoencoder (VAE)~\cite{kingma2013auto} for generating age-converted brain images, but often produce blurred results and lack auxiliary techniques for preserving individual identity. In contrast, \cite{xia2019consistent,xia2021learning} utilize a cGAN-based framework to generate aged brain images, although limited to synthesizing sliced 2D MRI scans. More recently, \cite{pombo2023equitable} proposes a conditional generative model employing diffeomorphic deformations to age-convert images based on the StarGAN~\cite{choi2018stargan} framework. However, these methods face challenges in adequately preserving the identity of the input subject.

\subsection{Identity Preservation in Age Transformation}
Age transformation studies, encompassing both face and brain aging, have evolved with a focus on preserving the identity of the input subject throughout the age conversion process. Conventional aging models attempt to maintain the subject's identity by employing a direct reconstruction loss at the image level. For instance, CAAE~\cite{zhang2017age} utilizes L2 loss between reference and generated images to preserve independent traits. However, this objective function presents a fundamental issue in generating blurry images due to incomplete gradient updates for sharpness. As an alternative approach, \cite{xia2019consistent,xia2021learning,he2019s2gan} adopt a pixel-wise L1 loss, resulting in sharper synthesized images. Nonetheless, they still face the limitation that the generated image does not guarantee perceptual quality.

In contrast, IPCGAN~\cite{wang2018face} proposes a perceptual loss that considers human perceptions in image similarity, utilizing a pre-trained AlexNet~\cite{krizhevsky2012imagenet} to aid identity conservation. Although this approach yields generated images closer to the reference image in perceptual criteria, it encounters limitations due to reliance on an auxiliary classifier pre-trained on ImageNet~\cite{deng2009imagenet} rather than being specifically trained for the desired domain of age transformation. As an alternative, recent studies on face aging~\cite{maeng2023age,li2023pluralistic} adopt face identity classifier~\cite{ubayashi2010archface} or disentangled representations containing the reference image's characteristics~\cite{li2020hierarchical,huang2021age}. While these methods may effectively preserve identity, their application to brain age transformation is hindered by the requirement for numerous longitudinal data.

In our study, we preserve the identity of the input subject by disentangling the representation extracted from an encoder into age-related and age-irrelevant features. The age-irrelevant features are then updated using a disentangled representation learning strategy. As a result, without necessitating longitudinal data or a pre-trained identity verification module, IdenBAT adeptly preserves identity while enabling precise age transformation to the target ages.

\begin{figure}[t]
    \centering
    {\includegraphics[width=0.65\textwidth]{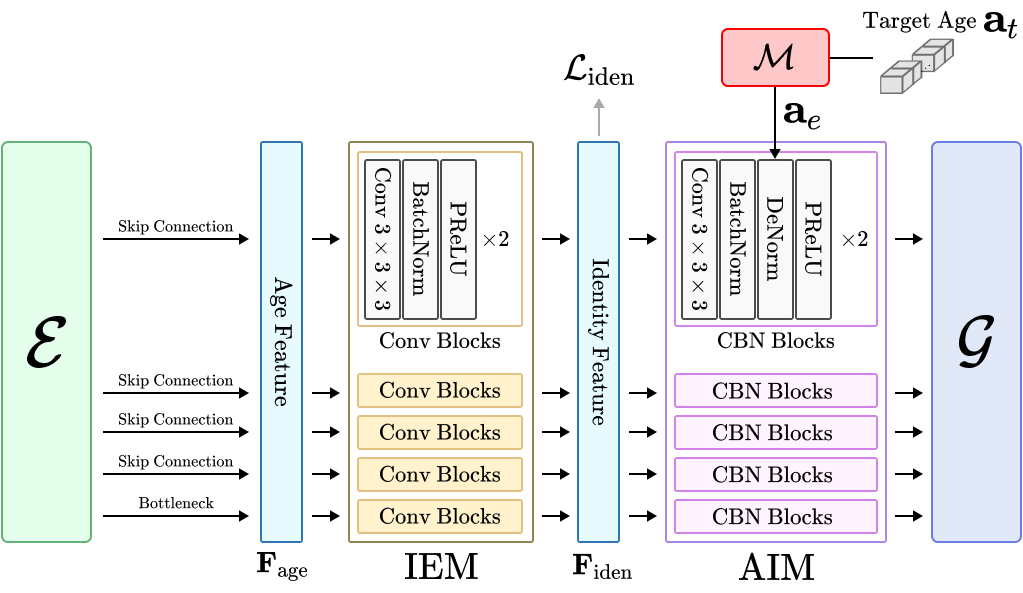}}
    \caption{Detailed architecture of IEM and AIM. DeNorm and CBN refer to the denormalization process and conditional batch normalization, respectively.}\label{fig:IEM}
\end{figure}

\section{Methods}
The primary objective of IdenBAT is to transform the image precisely to the target age while meticulously preserving the intrinsic identity of the input brain image. We design the identity extracting module (IEM) to disentangle intertwined semantic information extracted from the input image to realize this objective. Leveraging the isolated identity feature, we incorporate age conditions into the age injecting module (AIM) to synthesize age-converted images. Additionally, employing a conditional discriminator and age classifier, our aging synthesizer more accurately captures age changes and enhances visual quality. The overall framework of our proposed method is illustrated in Fig.~\ref{fig:architecture1}.

\subsection{Age Transformer}
Given an input image $\mathbf{X}\in\mathbb{R}^{H\times W\times D}$ and a randomly selected target age $\mathbf{a}_t$, our age transformer $\mathcal{T}$ aims to synthesize an age-transformed image $\hat{\mathbf{X}}\in\mathbb{R}^{H\times W\times D}$ as follows:
\begin{equation}
    \hat{\mathbf{X}}:= \mathcal{T}(\mathbf{X}, \mathbf{a}_t).
\end{equation}
The aged transformer, structured as a U-Net~\cite{ronneberger2015u}, comprises an encoder $\mathcal{E}$, a generator $\mathcal{G}$, IEM, and AIM. Fig.~\ref{fig:IEM} illustrates the detailed architecture of IEM and AIM.

The role of the encoder $\mathcal{E}$ is to extract entangled features that predominantly represent the personal age of the reference image, denoted as $\mathbf{F}_\text{age}=\{\mathbf{F}^1_\text{age},...,\mathbf{F}^L_\text{age}\}$, where $\{\mathbf{F}_{\text{age}}^l\}_{l=1}^{L}$ represents the feature maps of the $l$-th convolution blocks in the encoder $\mathcal{E}(\mathbf{X})$. Each feature map $\mathbf{F}_{\text{age}}^l\in\mathbb{R}^{H'\times W'\times D'}$ undergoes size reduction from $H\times W\times D$ through max pooling. The encoder is updated using an age prediction loss, with the current age of the input image serving as the ground truth. This ensures that the age feature $\mathbf{F}_\text{age}$ accurately reflects the subject's actual age in the image.

The age transformation process involves altering the condition while inherently preserving the individual's unique characteristics. To seamlessly maintain identity while incorporating the age condition in the age transformer $\mathcal{T}$, our objective is to disentangle the identity feature from the age feature. This disentanglement is essential to ensure that the model effectively captures the essence of age conversion without compromising the distinct identity of the subject. To achieve this, we extract the identity feature $\mathbf{F}_\text{iden}=\{\mathbf{F}^1_\text{iden},...,\mathbf{F}^L_\text{iden}\}$ from the entangled age feature $\mathbf{F}_\text{age}$ through the IEM, which will be elaborated on the following subsection.

We utilize a style transfer method for age conditioning in AIM to incorporate information about the target age into the identity feature. Here, $\text{AIM}=\{\text{AIM}^1,...,\text{AIM}^L\}$ consists of a series of conditional batch normalization (CBN)~\cite{de2017modulating} blocks of $l$-th layers, represented as:
\begin{equation}
    \text{AIM}^l=(\operatorname{Conv} \rightarrow \operatorname{BN} \rightarrow \operatorname{DN} \rightarrow \operatorname{PReLU}) \times 2.
\end{equation}
BN and DN represent the batch normalization and denormalization processes, respectively. The target age $\mathbf{a}_t$ is incorporated into $\mathbf{a}_e$ using a mapping network $\mathcal{M}$, comprising eight fully-connected layers with LeakyReLU activation function, in line with StyleGAN2~\cite{karras2020analyzing}. The embedded age condition is fused into the identity feature $\mathbf{F}_\text{iden}$ through CBN blocks. Specifically, the $l$-th identity feature 
$\mathbf{F}_\text{iden}^l$ undergoes batch normalization, standardizing the features batch-wise and mapping them into a specific normalized space. Subsequently, these normalized features undergo denormalization, wherein the target age representation $\mathbf{a}_e$ is incorporated by adjusting the mean and standard deviation parameters of the denormalization as:
\begin{equation}
    \mathbf{F}_\text{out}^l = \text{AIM}^l(\mathbf{F}_\text{iden}^l, \mathbf{a}_e).
\end{equation}
This technique allows for reintroducing age-specific characteristics into the normalized identity feature, tailoring it to reflect the desired age condition.

Finally, $\mathbf{F}_\text{out}=\{\mathbf{F}^1_\text{out},...,\mathbf{F}^L_\text{out}\}$ serve as input to the generator $\mathcal{G}$, yielding an age-transformed image 
$\hat{\mathbf{X}}$, with a residual connection to the input $\mathbf{X}$ as:
\begin{equation}
    \hat{\mathbf{X}} = \mathbf{X}+\mathcal{G}(\mathbf{F}_\text{out}).
\end{equation}
The residual connection ensures that the synthesized images retain substantial identities of the input, even in the early stages of training.

\subsection{Identity Extracting Module}
Recent advancements in brain age transformation~\cite{xia2019consistent,xia2021learning,pombo2023equitable} have seen approaches directly synthesizing age-transformed images based on features extracted from an encoder containing various attributes of the input image. However, this poses a substantial challenge in selectively preserving specific features, particularly those associated with the subject's identity. To address this challenge, we propose disentangling the identity feature $\mathbf{F}_\text{iden}$ from the age-related feature $\mathbf{F}_\text{age}$ as $\mathbf{F}_\text{iden}=\text{IEM}(\mathbf{F}_\text{age})$. Similar to the structure of AIM, the $l$-th layers of $\text{IEM}=\{\text{IEM}^1,...,\text{IEM}^L\}$ comprise two convolution blocks:
\begin{equation}
    \text{IEM}^l=(\operatorname{Conv} \rightarrow \operatorname{BN} \rightarrow \operatorname{PReLU}) \times 2.
\end{equation}
We employ two distinct loss functions based on cosine similarity measures to guide IEM in effectively extracting $\mathbf{F}_\text{iden}$ through convolution operations. First, considering that the input image $\mathbf{X}$ and age-converted image $\hat{\mathbf{X}}$ differ solely in the age condition, maximizing the cosine similarity between their identity features enables updating IEM to preserve the input subject’s identity. Subsequently, orthogonal projection between $\mathbf{F}_\text{age}$ and $\mathbf{F}_\text{iden}$ is performed by making their cosine similarity zero. These disentangled representation learning strategies enable the convolution blocks of IEM to effectively distinguish features intrinsic to the aging process from unrelated ones, thereby enhancing the model's ability to preserve identity information.

A primary concern when manipulating U-Net-based architecture bottleneck features is the potential information loss in skip connections. Specifically, suppose feature disentanglement is solely conducted at the last feature (\ie, bottleneck) of the encoder $\mathcal{E}$ without a similar process in the skip connections linked to the generator $\mathcal{G}$. In that case, there is a risk of information loss. To address this issue, we designed AIM and IEM to perform identity feature disentanglement and target age conjunction at each $l$-th skip connection and bottleneck as:
\begin{equation}
    \mathbf{F}_\text{out}^l = \text{AIM}^l(\text{IEM}^l(\mathbf{F}_\text{age}^l), \mathbf{a}_e).
\end{equation}
This scheme ensures that feature disentanglement and age injection are consistently applied across all levels of the skip connections, as depicted in Fig.~\ref{fig:IEM}.

\subsection{Conditional Discriminator}
Our model introduces a novel approach to brain age transformation by integrating target age conditions into the discriminator. This conditioning method substantially augments the discriminator's ability to authenticate the generated images and evaluate their alignment with the specified age conditions. To accomplish this, we have devised a conditional discriminator $\mathcal{D}$ that plays a crucial role in the adversarial learning process. The architecture of $\mathcal{D}$ draws inspiration from PatchGAN~\cite{isola2017image}, comprising a series of convolution blocks with a critical adaptation to incorporate age conditioning effectively. Specifically, we integrated CBN within the first convolution block of $\mathcal{D}$ to facilitate the injection of age conditions. Similar to the approach in AIM, the age condition $\mathbf{a}_t$ undergoes processing through the mapping network $\mathcal{M}$, followed by denormalization into the first feature of the discriminator, thus conveying the target age information. By employing this sophisticated conditioning mechanism, the discriminator can critically evaluate the age accuracy of the generated images, thereby enhancing the model's overall ability to produce age-authentic transformations.

\subsection{Objectives}
We outline the essential objective functions to optimize our proposed network, including age prediction, identity-extracting, cycle-consistency, reconstruction, and adversarial losses.

\subsubsection{Age Prediction Loss}
We leverage age prediction loss to extract the age representation from the input image and further facilitate the age transformer $\mathcal{T}$ in accurately converting the given image to the target age.

First, we minimize a Kullback–Leibler (KL) divergence loss function between the predicted probability of the encoder $\mathcal{E}$ and a Gaussian distribution, where the mean corresponds to the age of the input image $\mathbf{a}_i$ and the standard deviation is set to one following~\cite{peng2021accurate}:
\begin{equation}
\mathcal{L}_{\text{age}_1}=\mathcal{L}_\text{KL}(\mathcal{E}(\mathbf{X}),{\mathbf{a}_i}).
\end{equation}
Here, $\mathcal{L}_\text{KL}$ denotes the KL divergence loss. Once the encoder is updated to predict the age of the given image, we utilize the frozen encoder $\mathcal{E}^{*}$ to guide the age transformer $\mathcal{T}$ to accurately synthesize an age-converted image $\hat{\mathbf{X}}$ through another age prediction loss as:
\begin{equation}
\mathcal{L}_{\text{age}_2}=\mathcal{L}_\text{KL}(\mathcal{E}^{*}(\hat{\mathbf{X}}),{\mathbf{a}_t}),
\end{equation}
where $\hat{\mathbf{X}} = \mathcal{T}(\mathbf{X}, \mathbf{a}_t)$ and $\mathbf{a}_t$ denotes the target age condition.

\subsubsection{Identity-Extracting Loss}
To effectively disentangle the representation of identity from the input subject, we propose an identity-extracting loss that integrates both cosine similarity and orthogonality losses. Inspired by recent advancements in self-supervised learning methods~\cite{chen2021exploring,grill2020bootstrap}, we consider the age-converted image $\hat{\mathbf{X}}$ as an augmented view of the input image $\mathbf{X}$. By maximizing the cosine similarity between the identity features $\mathbf{F}_\text{iden}$ extracted from this pair, the IEM is trained to extract identity features seamlessly, independent of age information. Additionally, we employed an orthogonality loss between $\mathbf{F}_\text{age}$ and $\mathbf{F}_\text{iden}$ to fully disentangle identity information from the entangled representation $\mathbf{F}_\text{age}$. This is achieved by enforcing the cosine similarity between $\mathbf{F}_\text{age}$ and $\mathbf{F}_\text{iden}$ to be zero. The final identity-extracting loss, which combines the cosine similarity loss and the orthogonality loss, is defined as follows:
\begin{equation}\label{id}
\begin{aligned}
\mathcal{L}_\text{iden}=-\langle\mathbf{F}_\text{iden},\hat{\mathbf{F}}_\text{iden}\rangle+|\langle\mathbf{F}_\text{iden},\mathbf{F}_\text{age}\rangle|,
\end{aligned}
\end{equation}
where $\hat{\mathbf{F}}_\text{iden}$ represents the age-irrelevant feature of the age-converted image $\hat{\mathbf{X}}$ and $\langle\cdot\rangle$ denotes cosine similarity.

\subsubsection{Cycle-Consistency Loss}
To ensure identity preservation across age transformations within our model, we incorporate the concept of cycle-consistency loss. This loss mechanism plays a vital role in maintaining the core identity features of brain images as they undergo age progression and regression cycles. The process involves transforming an image of the current age $\mathbf{X}$ to its target age, denoted as $\hat{\mathbf{X}}=\mathcal{T}(\mathbf{X},\mathbf{a}_t)$, and subsequently reverting this age-transformed image to its original age using the age condition of input image $\mathbf{a}_i$. The cycle-consistency loss is then computed as the L1 norm between the original image and the twice-transformed image, as depicted below:
\begin{equation}
\begin{aligned}
\mathcal{L}_{\text{cyc}}=\|\mathbf{X}-\mathcal{T}(\hat{\mathbf{X}},\mathbf{a}_i)\|_1.
\end{aligned}
\end{equation}
This approach emphasizes the importance of maintaining the semantic integrity of brain images throughout the transformation process. It safeguards against potential issues such as mode collapse, where the model might learn to map multiple age-varied inputs to a single output representation.

\subsubsection{Reconstruction Loss}
We further incorporate a reconstruction loss to uphold the structural integrity of brain images throughout the age transformation process. This loss mechanism involves applying an L2 norm-based reconstruction loss between the input image $\mathbf{X}$ and the synthesized age-transformed image $\hat{\mathbf{X}}$, effectively quantifying the discrepancy in their structural details. This loss is augmented by a weighting strategy~\cite{maeng2023age} that considers the absolute age difference between the input and target ages, as follows:
\begin{equation}
\begin{aligned}
\mathcal{L}_{\text{rec}}=(\beta\cdot\cos(\pi\cdot\Delta_{age})+(1-\beta))\cdot\|\mathbf{X}-\mathcal{T}(\mathbf{X},\mathbf{a}_t)\|_2^2.\label{eq:rec}
\end{aligned}
\end{equation}
Here, the weight applied to the L2 norm is dynamically adjusted based on the age difference $\Delta_{age}$, which is computed as the percentage difference between the input age $\mathbf{a}_i$ and the target age $\mathbf{a}_t$, \ie, $\Delta_{age}=\frac{1}{r}|\mathbf{a}_i-\mathbf{a}_t|$. The decrement factor $\beta$ modulates the influence of age variation, thereby ensuring that a larger age difference between the input and target ages results in a lesser weight on the reconstruction loss.

\subsubsection{Adversarial Loss}
In the final stage of our training strategy, we implemented an adversarial loss to facilitate GAN training. Leveraging the previously discussed conditional discriminator, our model adheres to the least square GAN (LSGAN)~\cite{mao2017least} methodology, employing a mean squared error (MSE) loss for the GAN objective. This choice aims to enhance the age transformer's ability to generate age-converted images that faithfully reflect the desired age characteristics, thereby minimizing the distance from the discriminator's decision boundary. The loss functions for adversarial learning are defined as follows:
\begin{equation}
\begin{aligned}
\mathcal{L}_\text{advD}=\frac{1}{2} \mathbb{E}_{\Bar{\mathbf{X}}\sim P_{\boldsymbol{x}}}\left[(\mathcal{D}(\Bar{\mathbf{X}},\mathbf{a}_t)-1)^{2}\right]+\frac{1}{2} \mathbb{E}_{\hat{\mathbf{X}}\sim P_{\boldsymbol{x}}}\left[(\mathcal{D}(\hat{\mathbf{X}},\mathbf{a}_t))^{2}\right],
\end{aligned}
\end{equation}
\begin{equation}\label{advG}
\begin{aligned}
\mathcal{L}_\text{advG}= \mathbb{E}_{\hat{\mathbf{X}}\sim P_{\boldsymbol{x}}}\left[(\mathcal{D}(\hat{\mathbf{X}},\mathbf{a}_t)-1)^{2}\right],
\end{aligned}
\end{equation}
where $\Bar{\mathbf{X}}$ represents the randomly selected real samples corresponding to the target age, and $P_{\boldsymbol{x}}$ denotes data distribution.

The training of our model is conducted in an end-to-end manner, updating all objectives jointly to balance the contributions of each loss component:
\begin{equation}
\begin{aligned}
\min\lambda_\text{adv}\mathcal{L}_\text{advG}+\lambda_\text{age}(\mathcal{L}_{\text{age}_1}+\mathcal{L}_{\text{age}_2})
\\
+\lambda_\text{iden}\mathcal{L}_\text{iden}+\lambda_\text{cyc}\mathcal{L}_\text{cyc}+\lambda_\text{rec}\mathcal{L}_\text{rec},
\end{aligned}
\end{equation}
\begin{equation}
\begin{aligned}
\min\mathcal{L}_\text{advD},
\end{aligned}
\end{equation}
where $\lambda_{\text{adv}\backslash\text{age}\backslash\text{iden}\backslash\text{cyc}\backslash\text{rec}}$ represent the weighting hyperparameters of each loss. This comprehensiveness strategy ensures an effective learning process, optimizing for realistic age transformations while preserving the integrity of the brain images.

\begin{figure*}[t]
    \centering
    \includegraphics[scale=0.112]{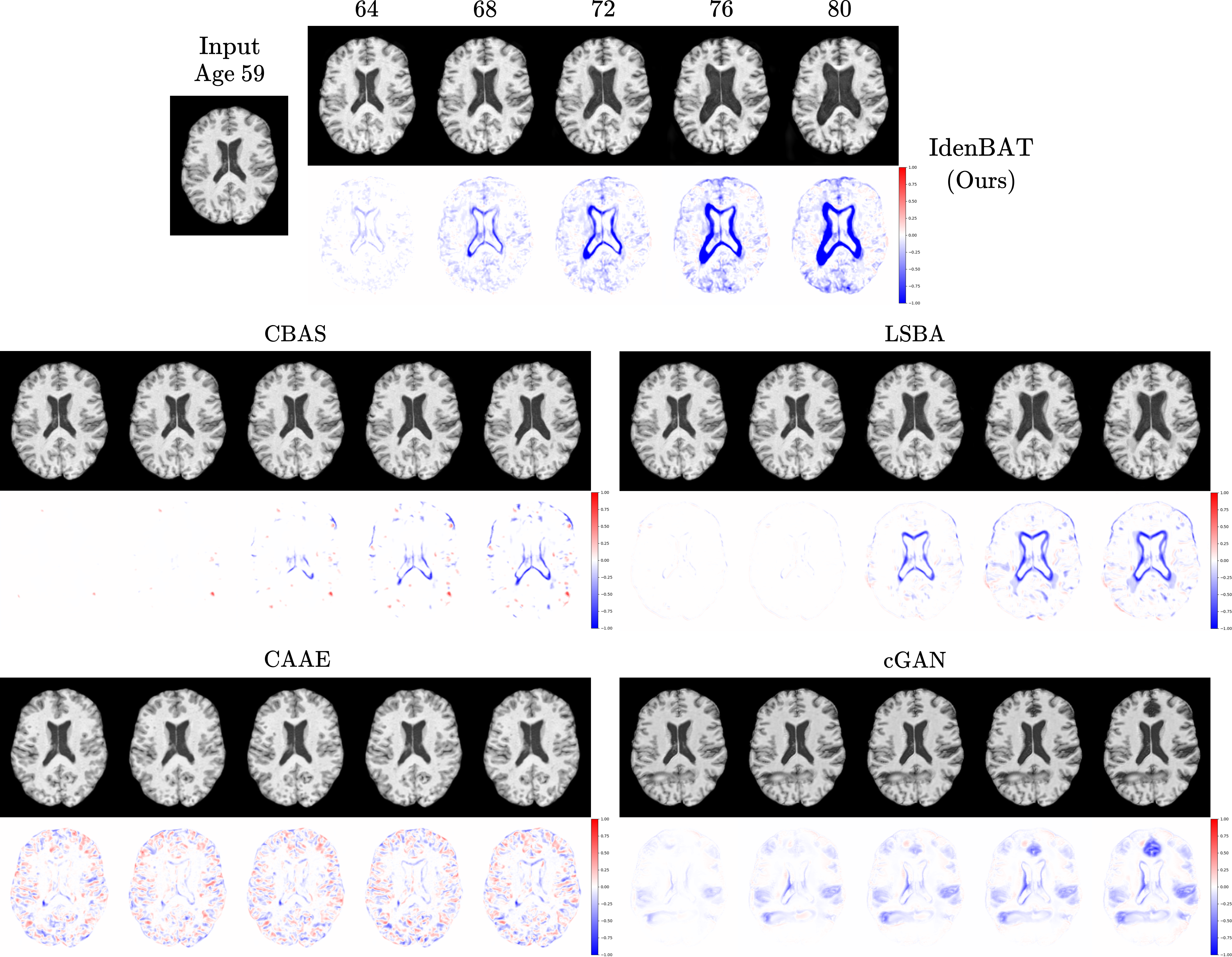}
    \caption{Qualitative comparison of brain aging with different ages on 2D brain images. Difference maps are synthesized by subtracting the input image from the age-converted images.}\label{fig:2d}
\end{figure*}

\section{Experiments}

\subsection{Datasets and Preprocessing}
We utilized the UK Biobank\setcounter{footnote}{0}\footnote{\url{https://www.ukbiobank.ac.uk}} dataset to validate the effectiveness of our proposed IdenBAT. This dataset comprises 49,123 T1 MRIs from individuals with healthy conditions. We selected 7,590 images from this dataset for both 2D and 3D training data, covering ages ranging from 48 to 80. We randomly sampled 230 images for each age group to mitigate potential class imbalance issues. Additionally, we employed full-size 3D images with a resolution of $193\times229\times193$. For the 2D images, we selected the 97th axial slice from the preprocessed images, corresponding to the central plane of the axial view. This resulted in 2D images with a resolution of $229\times193$, as per the methodology outlined in~\cite{xia2019consistent}. To assess the efficacy of our proposed framework comprehensively, we gathered longitudinal test data based on baseline subjects. This longitudinal dataset comprised 982 pairs of images with an age gap exceeding two years. Initially, we conducted neck removal from the raw brain images using the “robustfov” tool provided in the FMRIB Software Library (FSL)~\cite{jenkinson2012fsl}. Subsequently, we applied HD-BET~\cite{isensee2019automated} for brain extraction to remove non-brain tissue from the entire head image. The resulting skull-stripped images were then registered to the MNI152 template using the FLIRT tool for linear registration. Finally, we obtained preprocessed 3D MRI images with dimensions of $193\times 229\times193$, following the default parameters of the preprocessing toolkit.

\subsection{Implementation Details}
To illustrate the scalability of our proposed IdenBAT, we conducted experiments utilizing both 2D and 3D brain images. We implemented center cropping to reduce unnecessary computational costs, reducing the model’s input size to $176\times208\times176$. For the 2D data, we removed redundant background areas through cropping, resulting in images of $208\times176$ resolution. In both the 3D and 2D brain aging models, we employed the Adam optimizer~\cite{kingma2014adam} with a learning rate of $10^{-3}$ for the encoder, $5\times10^{-4}$ for the generator, AIM, and discriminator. We utilized a learning rate of $10^{-5}$ to update the mapping network and IEM. To mitigate the risk of overfitting, we utilized a StepLR scheduler with a step size of 30 and a gamma of 0.3. Additionally, we employed an augmentation strategy during training, including random flipping (horizontal flip for 2D and sagittal flip for 3D images) and Gaussian blurring for the encoder. The weights for the loss functions $\lambda_\text{adv}$, $\lambda_\text{age}$, $\lambda_\text{iden}$, $\lambda_\text{cyc}$, and $\lambda_\text{rec}$ were set to 1, 0.05, 1, 0.1, and 0.1, respectively. For the hyperparameter setting within $\delta_{\text{age}}$ for Eq. \eqref{eq:rec}, we defined the value of $r$ as 33. We trained our IdenBAT for 50 epochs with a batch size of 4 using 3D data and for 200 epochs with a batch size 64 using 2D data, utilizing two NVIDIA RTX A6000 GPUs.

\subsection{Baseline Models}
We conducted a comprehensive comparative analysis of our IdenBAT against several state-of-the-art brain age transformation models with official implementations. For the 2D brain age transformation task, we selected cGAN~\cite{mirza2014conditional}, CAAE~\cite{zhang2017age}, CBAS~\cite{xia2019consistent}, and LSAB~\cite{xia2021learning} as baseline models. Since cGAN utilizes its input as the noise vector, we employed a U-Net-based encoder for conditional image-to-image translation. However, it is important to note that these models were primarily designed for 2D images, which are slices of the original 3D brain images. Consequently, they did not produce satisfactory results when applied to 3D brain age transformation tasks. For the comparison involving full-size 3D brain age transformation, we included CounterSynth~\cite{pombo2023equitable} as the sole model. This choice reflects the substantial challenge of completely transforming brain age using full-size 3D models. We emphasize that modeling brain age transformation using full-size 3D data poses considerable difficulties and requires specialized techniques.

\begin{figure*}[t]
    \centering
    \includegraphics[scale=0.103]{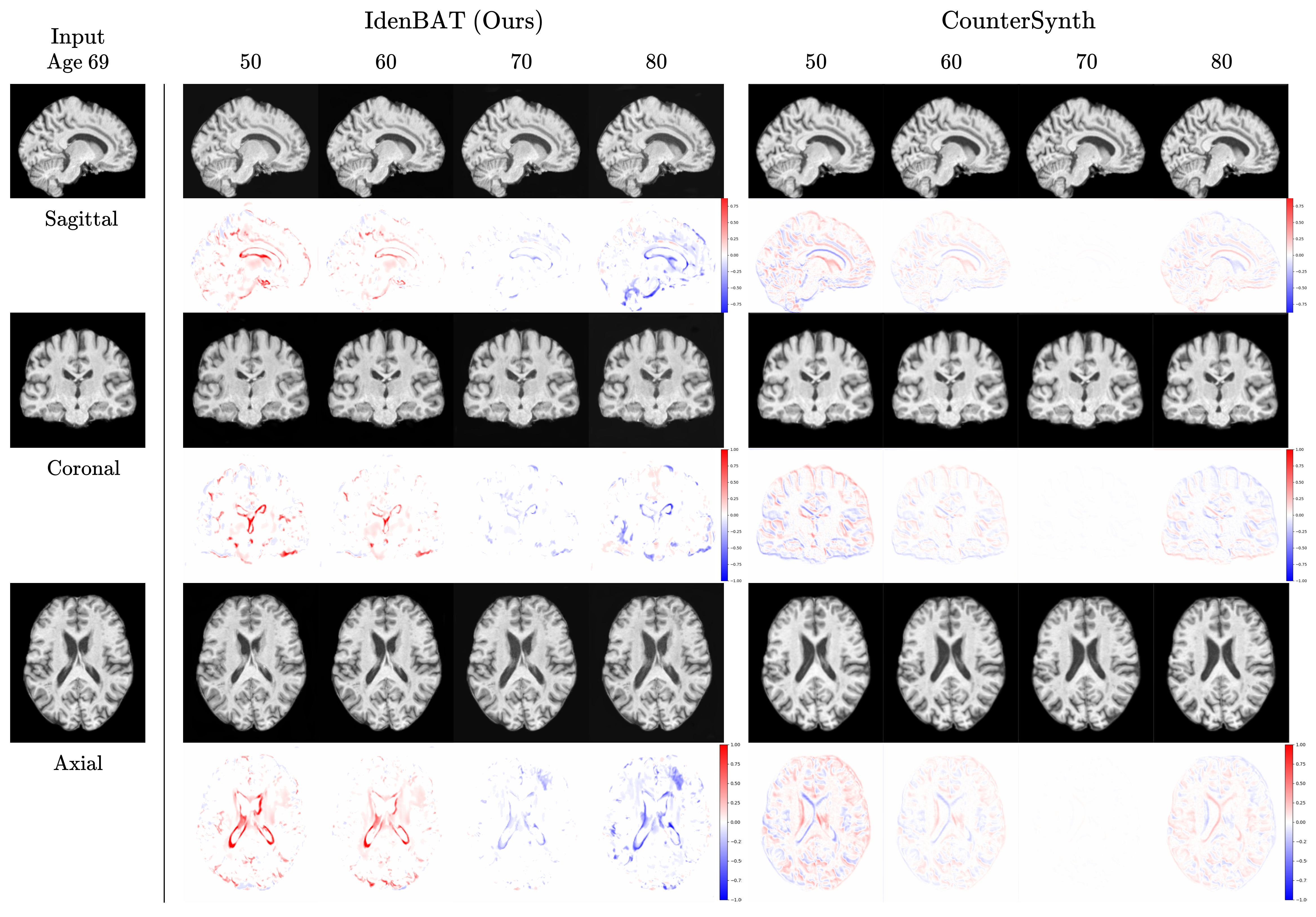}
    \caption{Qualitative comparison of brain aging across various ages on 3D brain images.}\label{fig:3d}
\end{figure*}

\begin{table*}[t!]
\centering
\caption{Quantitative evaluation results between brain aging evaluation and age transformation accuracy in different dimension settings (2D and 3D).}
\renewcommand{\arraystretch}{1.1}
\resizebox{\textwidth}{!}{%
\begin{tabular}{ccccc|cccccc}
\toprule
\multirow{1}{*}{\textbf{Dimension}} & \multirow{2}{*}{\textbf{Methods}} & \multicolumn{3}{c}{\textbf{Brain Aging Evaluation}} & \multicolumn{6}{c}{\textbf{Age Transformation Accuracy (PAD ($\downarrow$))}} \\
\textbf{(2D / 3D)} & & \textbf{PSNR ($\uparrow$)} & \textbf{SSIM ($\uparrow$)} & \textbf{MSE ($\downarrow$)} & \textbf{All} & \textbf{(48 - 54)} & \textbf{(55 - 61)} & \textbf{(62 - 68)} & \textbf{(69 - 74)} & \textbf{(75 - 80)} \\
\cmidrule(lr){1-1} \cmidrule(lr){2-2} \cmidrule(lr){3-5} \cmidrule(lr){6-11}
& cGAN~\cite{mirza2014conditional} & 23.835 & 0.905 & 0.0043 & 7.78 & 5.21 & 3.41 & 5.07 & 7.24 & 12.72 \\
 & CAAE~\cite{zhang2017age} & 18.743 & 0.660 & 0.0135 & 7.53 & 3.39 & 3.18 & 5.69 & 7.92 & 11.11 \\
\text{2D} & LSAB~\cite{xia2021learning} & 26.186 & 0.604 & 0.0092 & 5.39 & 4.30 & 4.00 & 5.02 & 5.05 & 6.73 \\
 & CBAS~\cite{xia2019consistent} & 26.171 & 0.790 & 0.0096 & 5.84 & 3.49 & 3.89 & 4.99 & 5.62 & 7.90 \\
 & \textbf{IdenBAT (Ours)} & \textbf{26.314} & \textbf{0.924} & \textbf{0.0026} & \textbf{3.52} & \textbf{3.27} & \textbf{3.05} & \textbf{3.56} & \textbf{3.14} & \textbf{4.72} \\
\midrule
\multirow{2}{*}{\text{3D}} & CounterSynth~\cite{pombo2023equitable} & 22.879 & 0.771 & 0.0062 & 9.06 & 13.15 & 7.45 & 4.77 & 7.28 & 12.96 \\
& \textbf{IdenBAT (Ours)} & \textbf{29.653} & \textbf{0.958} & \textbf{0.0020} & \textbf{4.68} & \textbf{5.24} & \textbf{3.47} & \textbf{4.26} & \textbf{5.27} & \textbf{5.35} \\
\bottomrule
\end{tabular}
}
\label{table:quantitative}
\end{table*}

\subsection{Evaluation Metrics}
For the quantitative evaluations of identity preservation and accurate age conversion, we employed the four primary metrics: peak signal-to-noise ratio (PSNR), structural similarity index (SSIM), mean squared error (MSE), and predicted age difference (PAD). 

Specifically, PSNR quantifies the difference between the original and generated images by calculating the logarithmic ratio of the peak signal power to the noise power. A higher PSNR value indicates better quality and closer resemblance to the original image. Unlike PSNR, which focuses on pixel-level differences, SSIM assesses the perceived quality by comparing local patterns of pixel intensities normalized for luminance and contrast. MSE provides a straightforward measure of error magnitude, with lower values indicating higher image similarity. Finally, PAD is a metric designed to evaluate the accuracy of the age transformation. We utilize a pretrained age classifier~\cite{peng2021accurate} to assess how closely the age-transformed images match the target age. By calculating the difference between the predicted age of the generated image and the target age, PAD provides a direct measure of the model's ability to transform age accurately.

\begin{figure}[t]
    \centering
    {\includegraphics[width=0.60\columnwidth]{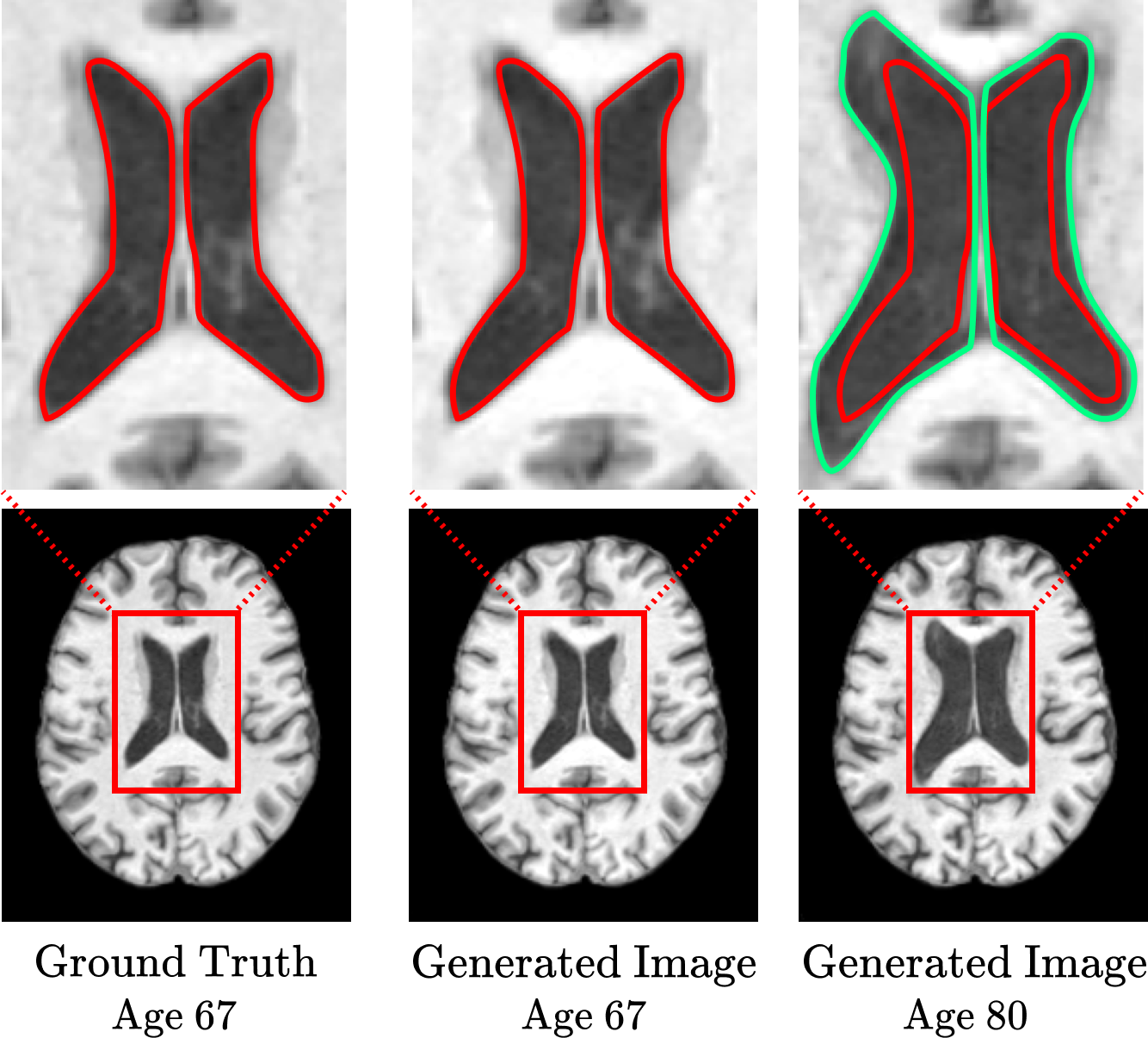}}
    \caption{Close-up of the ventricular contour of a subject.}\label{fig:close-up}
\end{figure}

\subsection{Qualitative Evaluation}
We utilized ground truth images to evaluate the accuracy of the age transformation process. For this purpose, longitudinal data from the UK Biobank was obtained, comprising MRI scans of the same subjects on two different occasions. We generated difference maps from the ground truth longitudinal data and those produced by various age transformation models, as depicted in Fig.~\ref{fig:gt}. Notably, the synthesized difference maps from our proposed IdenBAT, encompassing both 2D and 3D, closely resemble those of the ground truth (highlighted by the red box).
In contrast, while some comparative models (CBAS and LSBA) show only minor age changes for small age gaps (\eg, five years), others, like cGAN, synthesize artifacts. Meanwhile, CAAE and CounterSynth fail to preserve the subject's structural identity. These qualitative findings highlight that models producing images most similar to the second set of images from longitudinal data are more accurate in reflecting an individual's aging process, closely aligning with genuine brain aging. In essence, our proposed IdenBAT effectively preserves the unique characteristics of the input image while precisely transforming it to the target ages.

However, this experimental comparison using ground truth data has limitations in explaining transformations across large age gaps. To address this, we visualize the age-converted images and their difference maps over a wider age range in Fig.~\ref{fig:2d} for 2D and Fig.~\ref{fig:3d} for 3D. In Fig.~\ref{fig:2d}, a randomly selected 2D brain image of age 59 was transformed into ages ranging from 64 to 80. Odd and even rows represent the age-transformed images and the difference maps between the input and age-converted images. The results from our proposed IdenBAT demonstrate a smooth and continuous expansion in the size of the ventricles, cerebrospinal fluid (CSF), and sulci, consistent with the known normal brain aging path~\cite{sivera2019model,huizinga2018spatio}. In contrast, comparative methods struggle to depict clear and continuous aging or tend to produce meaningless artifacts. Furthermore, Fig.~\ref{fig:3d} offers a qualitative evaluation of 3D brain age transformation through sagittal, coronal, and axial slice images.
Specifically, a reference image at age 69 was randomly selected and transformed across ages 50 to 80. Similar to the 2D results, the size of the ventricles and CSF exhibits noticeable expansion with aging, contrasting with the comparative method, which scarcely preserves the subject's structural identity.

Lastly, for a more detailed comparison, we zoomed in on the ventricle areas of both the ground truth and generated images, as depicted in Fig.~\ref{fig:close-up}. The image synthesized with age 67 (middle) precisely matches the ventricular size in the ground truth (left), while the age-converted image to age 80 (right) shows an appropriately enlarged area. Through rigorous qualitative evaluation, we confirm that our proposed IdenBAT successfully performs age-appropriate transformations, simulating the aging process with a high level of identity preservation and accuracy.

\begin{figure}[t]
    \centering
    {\includegraphics[width=0.65\columnwidth]{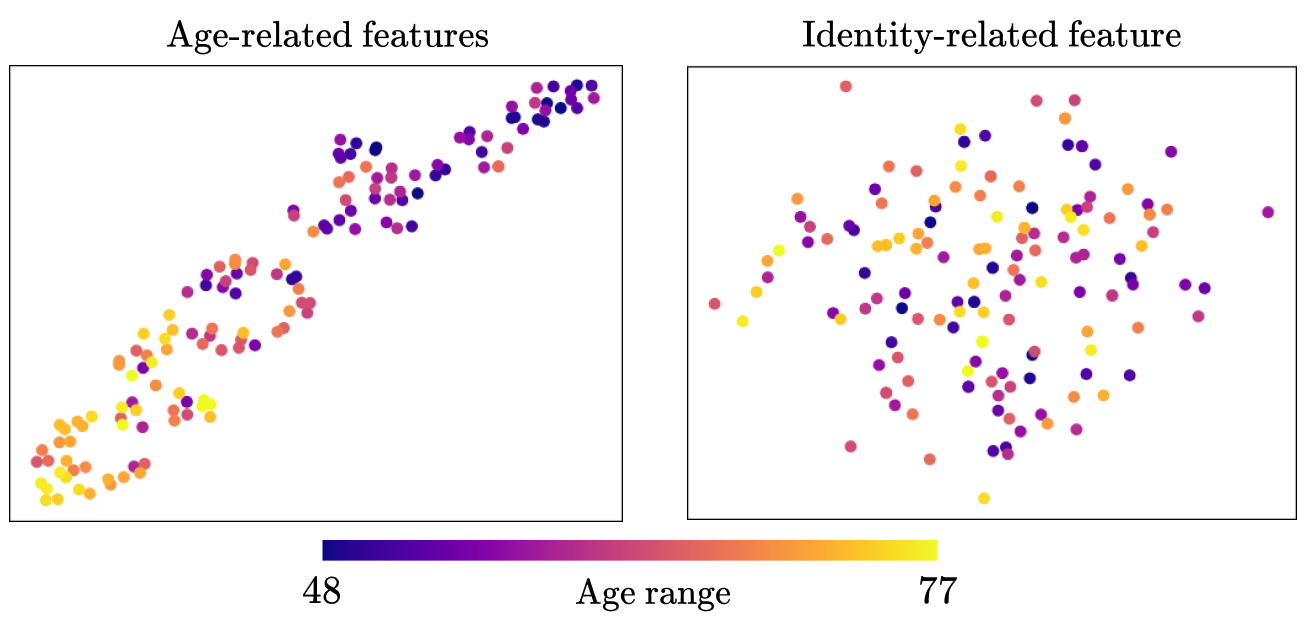}}
    \caption{t-SNE plots of age-related features (left) and identity-related features (right). Results of age-related features show continuous age representation. On the other hand, identity-related features are scattered across the test data due to their varying identities.}\label{fig:tsne}
\end{figure}

\begin{table}[t!]\centering\scriptsize\setlength{\tabcolsep}{5.1pt}
\renewcommand{\arraystretch}{1.15}
\caption{Quantitative evaluation for ablation studies. Case 1 denotes the model without using the identity-extracting loss. Case 2 and Case 3 represent ablating the cosine similarity loss and the orthogonality loss, respectively. Case 4 denotes the model with our proposed conditional discriminator replaced by a conventional discriminator.}
\centering
    \scalebox{1.2}{\begin{tabular}{ccccc}
    \toprule
    \textbf{Method} & \textbf{PSNR ($\uparrow$)} & \textbf{SSIM ($\uparrow$)} & \textbf{MSE ($\downarrow$)} & \textbf{PAD ($\downarrow$)} \\
    \cmidrule(lr){1-1} \cmidrule(lr){2-5}
    \text{Case 1}  & \text{24.875}   & \text{0.862}  & \text{0.0035}  & \text{4.27}  \\
    \text{Case 2}  & \text{25.006}   & \text{0.902}  & \text{0.0034}  & \text{4.15}  \\
    \text{Case 3}   & \text{25.091}   & \text{0.908}  & \text{0.0036}  & \text{3.60}  \\
    \text{Case 4}   & \text{25.934}   & \text{0.815}  & \text{0.0029}  & \text{3.86}  \\
    \textbf{Ours}       & \textbf{26.314} & \textbf{0.924} & \textbf{0.0026}  & \textbf{3.52} \\
    \bottomrule
    \end{tabular}}
\label{table:ablation}
\end{table}

\subsection{Quantitative Evaluation}
To further validate our model's performance, we utilized longitudinal test data for quantitative evaluation in Table~\ref{table:quantitative}. Initially, we transformed the test images $(t)$ into the target test images $(T)$ injecting the age information of the latter. Subsequently, we assessed the similarity between the synthesized images $(t \rightarrow T')$ and the corresponding ground truth images $(T)$ using PSNR, SSIM, and MSE metrics. By employing these metrics, we can effectively evaluate the degree of identity preservation in the brain age transformation models. High PSNR and SSIM values and low MSE values suggest that the generated images closely resemble the ground truth images in both structure and detail. This, in turn, indicates that the model successfully maintains the subject's identity during the age transformation process. As shown in Table~\ref{table:quantitative}, the results confirm the effectiveness of our IdenBAT in preserving the subject's identity and accurately matching the individual's aging trajectory.

Moreover, to evaluate the age accuracy of the transformed images, we computed PAD scores for the synthesized images, transforming all test images to ages ranging from 48 to 80 years. We adjusted the ages to older to ensure a fair comparison since most baseline models~\cite{zhang2017age,xia2019consistent,xia2021learning} were tailored for age progression. We trained a robust brain age prediction model~\cite{peng2021accurate} using 2D and 3D UK Biobank datasets, respectively, and utilized this model to estimate brain ages. As depicted in Table~\ref{table:quantitative}, our proposed IdenBAT surpassed baseline models across all age clusters and average scores.
These quantitative results lead us to conclude that our IdenBAT not only effectively preserves the characteristics of the input images but also accurately translates them to the designated target ages.

\begin{figure}[t]
    \centering
    {\includegraphics[width=0.65\columnwidth]{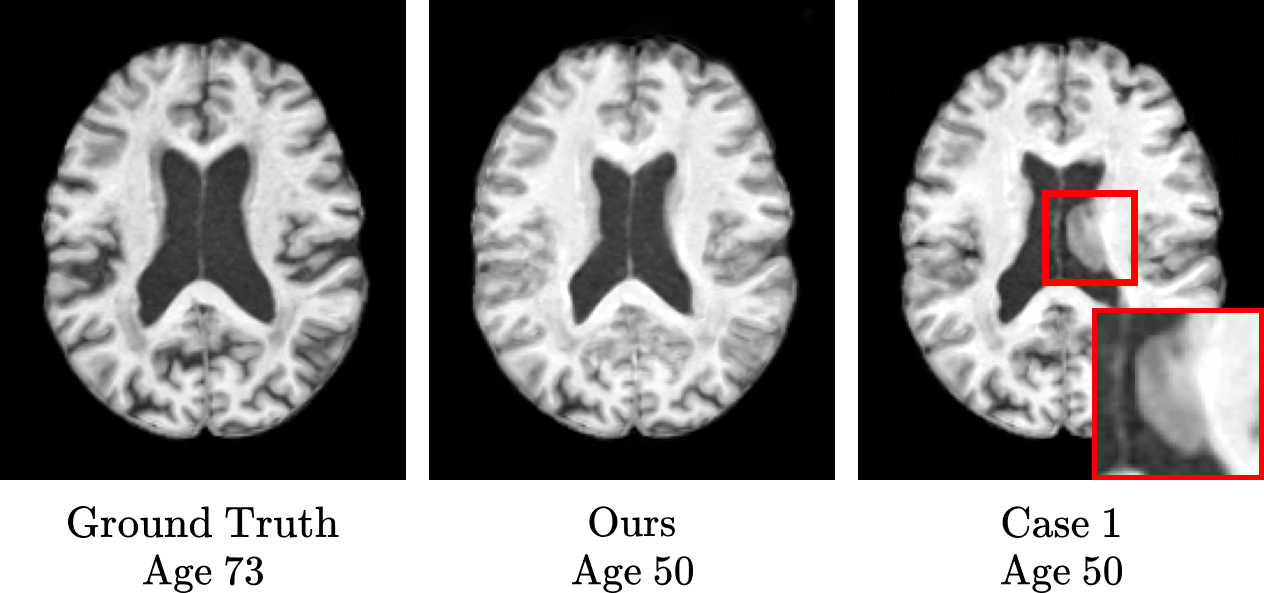}}
    \caption{Qualitative evaluation for the ablation study. The red box indicates the area that stands out when comparing experimental results.}\label{fig:ablation}
\end{figure}

\section{Model Interpretability and Ablation Study}
One of the primary contributions of IdenBAT is the disentanglement of age-related features $\mathbf{F}_\text{age}$ and identity-related features $\mathbf{F}_\text{iden}$. To validate the proper decomposition of these features, we randomly selected five subjects for each age in the test dataset and presented the t-SNE plots of their $\mathbf{F}_\text{age}$ and $\mathbf{F}_\text{iden}$ in Fig.~\ref{fig:tsne}. Interestingly, we observed that age-related features form a progressively coherent trajectory that aligns with the continuum of the aging process. Conversely, identity-related features are dispersed across the plot, indicating that the characteristics defining a person's identity remain consistent regardless of age. These visualized plots demonstrate our proposed IdenBAT effectively achieves the intended features disentanglement.

Furthermore, we conducted a series of ablation studies to verify the efficacy of specific components within our proposed method, as shown in Table~\ref{table:ablation}. Case 1 examines the impact of omitting the identity-extracting loss, thereby assessing its role in enhancing model performance. Cases 2 and 3 also explore the effects of removing the cosine similarity loss and the orthogonality loss from the identity-extracting loss, respectively. In Case 4, we investigate the significance of our conditioning mechanism within the discriminator by substituting it with a conventional cGAN~\cite{mirza2014conditional} approach. We observed that Case 1, which lacked all components of the identity-extracting scheme, demonstrated the lowest performance across all evaluation metrics. The results of Cases 2 and 3 also showed inferior scores compared to our complete model. The outcomes of Case 4 reveal the effectiveness of our proposed conditioning mechanism within the discriminator.
We depicted a qualitative result through the ablation study, as shown in Fig.~\ref{fig:ablation}. In the figure, it is evident that Case 1 struggles to maintain the structural identity of the ventricles, leading to undesirable shrinkage in areas that should remain unchanged (see red boxes). Consequently, we posit that these findings provide compelling evidence of the efficacy of our proposed IdenBAT in preserving the inherent identity while accurately converting ages to the desired target ages.

\section{Discussion}
In this work, we presented a brain age transformation model that enables precise age conversion of brain images while preserving individual identity. The clinical significance and contributions of our proposed IdenBAT offer substantial benefits in practical and medical applications.

First, IdenBAT opens the chance for clinicians to have an invaluable tool for visualizing and quantifying the brain aging process. By simulating the brain at various ages while preserving the identity of specific subjects, our model offers the opportunity to conduct highly reliable longitudinal studies that capture accurate morphological changes. A notable contribution of our IdenBAT is its ability to perform age transformations on full-size 3D MRI data. This is particularly significant in clinical settings as full-size 3D images provide a more comprehensive view of brain structures than 2D or cropped 3D images. Clinicians might leverage this capability to monitor more precise morphological characteristics within the aging process. Second, the characteristic of IdenBAT that does not require longitudinal data in training presents a significant advantage in clinical applications since collecting longitudinal MRI data is time-consuming and expensive. Also, reducing dependency on longitudinal data suggests that existing cross-sectional data can be utilized more effectively, potentially accelerating research and clinical studies. The last important aspect is the potential use for imputation in longitudinal datasets. In cases where some time points of the longitudinal data are incomplete or missing, our proposed IdenBAT could simulate plausible age-transformed images, filling in the gaps and enabling more comprehensive analyses.

Our proposed method also has avenues for improvement. IdenBAT can be adapted for use with other types of medical domains, such as T2-weighted sMRI and functional MRI (fMRI), or different neurodegenerative conditions. This adaptability could open new opportunities for the research community and clinicians to explore various aspects of brain structure and function across different imaging modalities and disease conditions.

\section{Conclusion}
This work proposed IdenBAT, a new framework for age transformation in the brain that focuses on preserving individual identity through disentangled representation learning. Compared to previous methods that primarily undertake identity preservation at the image level, the IdenBAT considered image-level and feature-level viewpoints, enabling the model to maintain personal characteristics. With age condition injection via style transfer mechanism, the IdenBAT enhanced the method's ability to model the complexities of the aging process. By further integrating cosine similarity and orthogonality objective functions, IdenBAT facilitated the extraction of an isolated representation of identity features from the input subject. Building on such advantages and through quantitative and qualitative evaluations, we demonstrated that the proposed method outperformed existing approaches that maintain individual identity during age alterations, providing a robust solution for accurately modeling the aging process in the brain. In conclusion, we believe that this advancement holds significant potential for applications in medical research and personalized aging treatment, where preserving individuals' unique identities is crucial.

\printcredits

\section*{Declaration of competing interest}
The author(s) declared no potential conflicts of interest with respect to the research, authorship, and/or publication of this article.

\section*{Acknowledgements}
This work was supported by Institute of Information \& communications Technology Planning \& Evaluation (IITP) grant funded by the Korea government (MSIT) No. RS-2019-II190079 (Artificial Intelligence Graduate School Program(Korea University)) and No. RS-2022-II220959 ((Part 2) Few-Shot Learning of Causal Inference in Vision and Language for Decision Making).

\bibliographystyle{cas-model2-names}

\bibliography{cas-refs}

\end{document}